\documentclass[preprint,prd,amssymb,amsmath,nobibnotes,nofootinbib]{revtex4}
\usepackage{amsfonts}
\usepackage{revsymb}
\usepackage{graphicx,epsfig}
\usepackage{placeins}

\begin{document}
\title {Construction of the second-order gravitational
 perturbations produced by a compact object}
\author{Eran Rosenthal}
\affiliation{
Department of Physics, University of Guelph, Guelph, Ontario  N1G 2W1,
Canada}

\date{\today}

\begin{abstract}

Accurate calculation of the gradual inspiral motion 
in an extreme mass-ratio binary system,  
in which a compact-object inspirals towards a supermassive black-hole
requires calculation of the interaction between 
the compact-object and the gravitational perturbations that it induces. 
These metric perturbations satisfy linear
partial differential equations on a curved background spacetime induced by 
the supermassive black-hole.
At the point particle limit the second-order perturbations equations 
have source terms that diverge as $r^{-4}$, where $r$ is the 
distance from the particle. This 
singular behavior renders the standard retarded solutions of these equations 
ill-defined. Here we resolve this problem and 
construct well-defined and physically meaningful
 solutions to these equations.
We recently presented an outline of this resolution \cite{eran1}. 
Here we provide the full details of this analysis.
These second-order solutions are important for practical calculations:
the planned gravitational-wave detector LISA requires 
preparation of waveform templates for the potential gravitational-waves. 
Construction of templates with desired accuracy for extreme mass-ratio
binaries requires accurate calculation of the inspiral motion including 
the interaction with the second-order gravitational perturbations. 
\end{abstract}

\maketitle

\section{Introduction}

Consider a binary system composed of a small compact-object with mass $\mu$ 
(e.g. a neutron star or a stellar-mass black-hole) that 
inspirals towards a supermassive black-hole with a mass $M$. 
Such extreme mass-ratio binaries (e.g., $M/\mu= 10^{5}$) 
are valuable sources for gravitational waves (GW) that could be detected 
by the planned laser interferometer space antenna (LISA) \cite{LISA}.
To detect these binaries and determine their parameters  
using matched-filtering data-analysis  
techniques one has to prepare gravitational  
waveform templates for the expected GW. 
An important part in the calculation  of the templates
 is keeping track of the 
GW phase. Successful determination of the binary parameters using  
matched-filtering techniques often requires one 
to prepare templates with a phase 
error of less than one-cycle over a year of inspiral \cite{BC}.
Calculating  waveform templates to this accuracy 
is a challenging task since 
 a waveform from a year of inspiral   
may contain $10^5$ wave-cycles \cite{TC}.

To carry out this calculation one is required to 
calculate the compact-object inspiral trajectory.
By virtue of the smallness of the mass ratio $\mu/M$
one may use perturbations analysis to 
simplify this calculation.
In this analysis the full spacetime metric  
is represented as a sum of a background metric -- 
induced by the supermassive black-hole, and a sequence of 
perturbations -- induced by the compact-object.
The object's trajectory is also treated with perturbations
techniques.
At leading order of this approximation 
the object's trajectory was found to be a geodesic 
in the background geometry (see e.g., \cite{DEATH}). 
At higher orders, the interaction between the object 
and its own gravitational field 
gives rise to a gravitational self-force that acts on the object. 
The leading order effect of the self-force originates  
from interaction between the object and its own 
first-order gravitational 
perturbations that are linear in $\mu$.
This first-order self-force (that scales like $\mu^2$) 
induces an acceleration of order $\mu$ for the object's trajectory.
In the case of a vacuum background geometry, 
formal and general expression for 
this first-order gravitational self-force  
was derived by Mino, Sasaki, and Tanaka \cite{MST},  
 and independently by Quinn and Wald \cite{QW} using a different method.
Later, practical  
methods to calculate this self-force were developed by several authors 
 \cite{PP,BO1,BMNOS,BO2,BO3,LOUSTO,BL,PD,HIKIDAETAL},
see also \cite{MINO,ENNAETAL} for a different approach to this problem. 
The next order corrections to the object's trajectory 
originate from the interaction between the object 
and its own  second-order gravitational perturbations (that are quadratic in $\mu$). 
Higher-order corrections will not be considered in this 
article.  

We shall now estimate the effect  
  of the gravitational self-force
 (more accurately the dissipative part
of the gravitational self-force, see below.)
on the accumulated phase of the emitted GW.
This will allow us to determine how many terms should be retained 
in the perturbations expansion (see also \cite{BURKO,POISSON,DETWEILER}). 
For simplicity consider a compact object which inspirals  
 between two circular orbits in a 
strong field region of a Schwarzschild black hole.
As the object inspirals towards the black hole its
 orbital frequency slowly changes from 
its value at an initial time. 
This shift in the orbital frequency is approximated by 
$\dot{\omega}t$, where
 $\dot{\omega}\equiv\frac{d\omega}{dt}$, 
and $t$ denotes the elapsed time (from initial time) 
in Schwarzschild coordinates.
Let $\Delta\phi$ denote 
the part of the phase shift of the GW (between two fixed times)
which is induced by the shift in the orbital frequency.
Since the GW frequency is proportional to the 
orbital frequency we 
find that after an inspiral time $\Delta t_{ins}$, the phase shift 
$\Delta\phi$ is approximately proportional to 
$\Delta t^2_{ins} \dot{\omega}$.
Let us find how the quantities in this expression scale with $\mu$.
The inspiral time $\Delta t_{ins}$ scales like 
$\Delta{E}\dot{E}^{-1}$,
here $E$ denotes 
the particle's energy per unit mass, 
$\Delta{E}$ is the energy difference between initial and final
 circular orbits, and $\dot{E}\equiv\frac{dE}{dt}$. 
The first-order gravitational self-force
 produces the leading term in an expansion of  
$\dot{E}$, which we denote $\dot{E}_1$. Since 
 $\dot{E}_1$ scales like $\mu$ we find that $\Delta t_{ins}$
scales like $M^2 \mu^{-1}$.
Turning now to $\dot{\omega}$, we write this quantity as 
$\dot{\omega}=\frac{d\omega}{dE}\dot{E}$.
At the leading order $\frac{d\omega}{dE}$ is independent of $\mu$ 
--- it is obtained from the equations 
describing a circular geodesic worldline. At higher orders 
 the conservative part of the self-force will produce
a correction to this geodesic orbit. Since we focus on the 
contribution to the phase coming from the dissipative part of the self-force
(i.e. the part of the self-force responsible for a non-zero $\dot E$.)
 we ignore the conservative corrections. 
Denoting the leading term in an expansion of $\dot{\omega}$
with  $\dot{\omega}_1$, and recalling that 
$\dot{E}_1$ scales like $\mu$ we  find that 
 $\dot{\omega}_1$ is of $O(\mu M^{-3})$.
Combining the expressions for $\Delta t_{ins}$ and  $\dot{\omega}_1$ 
we find that the first-order self-force produces 
a phase shift $\Delta\phi$ of order $\Delta t^2_{ins} \dot{\omega}_1=O(M/\mu)$.
The second-order 
self-force gives rise to second-order terms in the expansions of $\dot{E}$, and 
 $\dot{\omega}$. These terms are 
denoted here $\dot{E}_2$ and $\dot{\omega}_2$, respectively.
Since  $\dot{E}_2$ scales like $\mu^2$ we find that  $\dot{\omega}_2$ 
 is $O(\mu^2 M^{-4})$. 
After $\Delta t_{ins}$ the term $\dot{\omega}_2$ will produce  
a phase shift of order $\Delta t^2_{ins} \dot{\omega}_2=O[(M/\mu)^0]$.
Therefore, a calculation of $\Delta\phi$ to the desired 
accuracy of order $(M/\mu)^0$ (needed for LISA data analysis)
requires the calculation of the compact-object interaction with its 
own second-order metric perturbations.

The goal of constructing
 long waveform templates (e.g. for one year of inspiral)
which do not deviate by more than one-cycle from the 
true GW provides practical motivation for the study
of the second-order metric perturbations in this article.
Moreover, construction of second-order metric perturbations 
allows one to extend the applicability of the perturbation analysis to
binary systems with smaller $M/\mu$ mass-ratios. 
This study can also shed light 
on the problem of waveform construction.
Suppose that we attempt to construct a waveform 
for an inspiraling compact object (including the correct leading term for $\Delta \phi$) 
by using the following procedure.
First, we calculate a corrected worldline 
by including the contributions coming from the first-order self-force. 
Then we substitute this worldline into the expression of the source term
in the first-order perturbations wave-equation, and finally we construct 
a waveform by solving this equation. 
Here there is a subtlety, since the first-order
gravitational self-force is a gauge dependent quantity \cite{BOGAUGE}
(e.g. one may set it to zero by an appropriate choice of gauge, see below).
Therefore, by invoking a first-order gauge transformation 
 we can change the path of the corrected worldline. 
A waveform constructed from this new
corrected worldline may not encode the correct gauge invariant information. 
This 
argument reveals that for some gauge choices
 the above procedure does not 
provide us with a  waveform that include the correct leading term for $\Delta \phi$.
For these gauge choices it is reasonable to expect that the correct waveform 
could be obtained by including the second-order gravitational
 perturbations in the waveform calculation. 
In this this way all the contributions to the waveform which scale
like $\mu^2$ are being included in the calculation.

To construct the metric perturbations produced 
by a compact-object it is useful to consider the 
point particle limit -- where the dimensions of the compact object 
approach zero (below we give  more precise definitions of  the compact object and
the limiting process that we use).
In this limit the first-order metric perturbations
in the Lorenz gauge satisfy a wave equation with a delta function
source term (see e.g. \cite{PR}). 
It is well known that (certain components of) the retarded solution 
 of this wave equation diverges as $r^{-1}$ as the   
 worldline of the object is approached, 
where  $r$ denotes the spatial distance from the object. 
The construction of the second-order perturbations is more difficult.
In the limit the second-order metric perturbation equation away 
from the compact object take the following schematic form
\begin{equation}\label{2ndgrav} 
D[h^{(2)}]=\nabla h^{(1)}\nabla h^{(1)} \,\&\, h^{(1)}\nabla\nabla h^{(1)}\,. 
\end{equation} 
Here $\mu h^{(1)}$ and $\mu^2 h^{(2)}$ 
denote the first-order metric perturbations
and the second-order metric perturbations, respectively.
$D$ denotes a certain linear partial differential operator, 
$\nabla$ schematically denotes the covariant derivative with respect 
 to the background metric, and $\&$ denotes ``and terms of the form...''. 
Since  $h^{(1)}$ diverge as  $r^{-1}$ we find that the 
source term of Eq. (\ref{2ndgrav}) diverges as $r^{-4}$. 
One might naively attempt to construct a standard retarded
 solution to Eq. (\ref{2ndgrav}), by  
imposing Lorenz gauge conditions on $h^{(2)}$ and then formally integrating  
the singular source term with the corresponding retarded Green's function.
However, the resultant integral turns out to be  ill-defined, 
in fact it diverges at every point in spacetime. To 
see this notice that the invariant four-dimensional 
volume element scales like $r^2$ while the source term which is 
being integrated diverges as $r^{-4}$.

In this article we develop a regularization method
for the construction of  well-defined and physically meaningful solutions to
 Eq. (\ref{2ndgrav}).  A similar problem in a scalar toy-model
was recently studied \cite{scalar}. 
An outline of the resolution presented in this article was recently published \cite{eran1}.
Here we provide the complete details of this analysis, including derivations of 
results which were mentioned without derivations in \cite{eran1}. In addition  
we provide a prescription for the construction of Fermi-gauge which 
was only briefly mentioned in \cite{eran1}. 

For simplicity suppose that the small compact object is 
 a Schwarzschild black-hole. We consider the black hole to be
``small'' compared  to the  length scales that characterize Riemann curvature
 tensor of the vacuum background geometry (e.g. a stellar-mass  Schwarzschild black-hole in 
a strong field region of a background geometry induced by a supermassive Kerr black-hole).
 Denoting these length scales with  $\{{\cal R}_i\}$ we express our restriction 
as $\mu\ll\cal{R}$, where  ${\cal R}=\min\{{\cal R}_i\}$.
The presence of length scales with different orders 
of magnitude allows one to analyze this problem using the method 
of matched asymptotic expansions (see e.g. \cite{TH,PR}).
In this method one employs different approximation methods to calculate the
 metric in different overlapping regions of spacetime,
 where each approximation method is adapted to a particular subset of spacetime. 
Later, one matches the various metrics in these overlapping regions, 
and thereby obtain a complete approximate solution to Einstein's field equations.
In this article we shall consider the following decomposition of spacetime into 
two overlapping regions. Let $r$ be a meaningful notion 
of spatial distance, we define the internal-zone to lie within a worldtube 
which surrounds the black-hole and extends out to $r=r_I(\cal{R})$
 such that $r_I\ll\cal{R}$, and define the external-zone to lie 
 outside another worldtube $r=r_E(\mu)$, such that $\mu\ll r_E$. We denote the interior 
of this inner worldtube with $S$.
Since $\mu\ll\cal{R}$ we may choose $r_E$ to be smaller than $r_I$ such 
that there is an overlap between the above mentioned regions in $r_E<r<r_I$.
We shall refer to this overlap region as the buffer-zone 
(in the buffer-zone $r$ can be of order $\sqrt{\mu \cal{R}}$). 

This article is organized as follows: 
First in Sec. \ref{external} we discuss the perturbative approximation method 
to Einstein's field equations in the external-zone; 
in Sec. \ref{fop} we employ this approximation and construct 
the well known first-order metric perturbations;
in Sec. \ref{sop} we discuss the construction of the second-order 
 metric perturbations in the external-zone;
 in Sec. \ref{internal} we complete the construction of the 
physical second-order perturbations by matching the second-order external-zone
 solution to a solution in the internal-zone; finally 
Sec. \ref{conclusion} provides conclusions.

\section{Approximation in the External-zone}\label{external}

In the external-zone the spacetime geometry is dominated 
by the background geometry. Therefore, 
 it is convenient to decompose the full spacetime metric $g^{full}_{\mu\nu}$
into a background metric $g_{\mu\nu}$, 
 and perturbations  $\delta g_{\mu\nu}$ that are induced by the
 small black hole, reading
\begin{equation}\label{metricdec}
g^{full}_{\mu\nu}(x)=g_{\mu\nu}(x)+\delta g_{\mu\nu}(x)\,.
\end{equation}
Throughout this paper we use 
the background metric $g_{\mu\nu}$ to raise and lower tensor indices 
and to evaluate covariant derivatives. 
We expand $\delta g_{\mu\nu}$ in an asymptotic series, reading 
\begin{equation}\label{metricexp}
\delta g_{\mu\nu}(x)=\mu h^{(1)}_{\mu\nu}(x)+\mu^2 h^{(2)}_{\mu\nu}(x)+O(\mu^3)\,.
\end{equation}
Here the perturbations $\{h^{(i)}_{\mu\nu}\}$ are independent of $\mu$.

We shall now substitute the asymptotic expansion of $g^{full}_{\mu\nu}$ 
into Einstein's field equations and obtain linear 
partial differential equations for the first-order and second-order gravitational perturbations
 $h^{(1)}_{\mu\nu}$ and $h^{(2)}_{\mu\nu}$, respectively. 
We assume that full spacetime metric satisfies 
Einstein's field equations in vacuum, reading
\begin{equation}\label{Einsteineq}
R^{full}_{\mu\nu}=0\,.
\end{equation}
Here $R^{full}_{\mu\nu}$ is Ricci tensor of the full spacetime.
Substituting decomposition (\ref{metricdec})
 into Ricci tensor we obtain the following formal expansion
\footnote{For brevity we shall omit the tensors indices inside 
the squared brackets of the differential operators.}
\begin{equation}\label{riccidec}
R^{full}_{\mu\nu}=R_{\mu\nu}^{(0)}+R_{\mu\nu}^{(L)}[\delta g]+R_{\mu\nu}^{(Q)}[\delta g]+O(\delta g^3)\,.
\end {equation}
Here the superscripts denote the 
type of dependence on $\delta g _{\mu\nu}$:
$(0)$ --  no dependence on $\delta g_{\mu\nu}$
, $(L)$ --  linear dependence on $\delta g_{\mu\nu}$ 
and $(Q)$ -- quadratic dependence on $\delta g_{\mu\nu}$. 
Substituting Eq. (\ref{riccidec}) into Einstein equations (\ref{Einsteineq}) 
and using  expansion (\ref{metricexp}) we obtain 
\begin{eqnarray}\label{backgreq}
R_{\mu\nu}^{(0)}=0 \ \ \ \ \ ,&x\not\in S& \ \ \ \ \ ,\mu^0 \\\label{fstordreq}
R_{\mu\nu}^{(L)}[h^{(1)}]=0\ \ \ \ \ ,&x\not\in S& \ \ \ \ \ ,\mu^1 \\\label{scndordreq}
R_{\mu\nu}^{(L)}[h^{(2)}]=-R_{\mu\nu}^{(Q)}[h^{(1)}]\ \ \ \ \ ,&x\not\in S& \ \ \ \ \ ,\mu^2\,.
\end{eqnarray}
Note that Eq. (\ref{backgreq}) is an
 equation for the background
metric $g_{\mu\nu}$. This metric satisfies Einstein's field 
equations in the absence of the small black hole. We can therefore 
omit the restriction $x \not\in S$ from Eq. (\ref{backgreq}).

To further simplify the calculation 
it is useful to consider the limit $\mu\rightarrow 0$ of the series (\ref{metricexp}).
Notice that by definition  $h^{(1)}_{\mu\nu}$ and $h^{(2)}_{\mu\nu}$ do not  
depend on $\mu$ and therefore the form 
Eqs. (\ref{fstordreq},\ref{scndordreq}) is not  
affected by this limit. 
However, the domain of validity of these equations 
is in fact expanded as $\mu \rightarrow 0$.  
At this limit we let $r_E(\mu)$ approach zero 
 and Eqs. (\ref{fstordreq},\ref{scndordreq})
become valid throughout the entire background spacetime 
 excluding a timelike worldline $z(\tau)$, 
 where $\tau$ denotes proper time with respect to the background metric.
At this limit Eqs. (\ref{fstordreq},\ref{scndordreq}) take the form of
\begin{eqnarray}\label{newfst}
R_{\mu\nu}^{(L)}[h^{(1)}]=0&\,\,\,\,\,,x \not\in z(\tau)\,, \\ \label{newsnd}
R_{\mu\nu}^{(L)}[h^{(2)}]=-R_{\mu\nu}^{(Q)}[h^{(1)}]&\,\,\,\,\,,x \not\in z(\tau)
\,.
\end{eqnarray}

As they stand Eqs. (\ref{newfst},\ref{newsnd}) contain insufficient
 information about the physical properties of the  
sources that induce the perturbations. 
To obtain unique solutions to these equations,  
we must provide additional information about these sources.
The gravitational 
perturbations $h^{(1)}_{\mu\nu}$ and $h^{(2)}_{\mu\nu}$ are induced by a 
Schwarzschild black-hole, and therefore their properties 
on the worldline are determined from the physical properties of this 
source. 
As we show below these properties can be communicated to the external-zone 
 by specifying a set of divergent boundary conditions 
as $x\rightarrow z(\tau)$. Once these divergent boundary conditions
are specified, a physical solution (defined below) to the perturbation equations 
 (\ref{newfst},\ref{newsnd}) is uniquely determined.
In Sec. \ref{internal} below
we obtain the desired divergent boundary conditions
 for Eq. (\ref{newsnd}) from the corresponding internal-zone solution. 
For Eq.  (\ref{newfst}) D'Eath has shown \cite{DEATH}
that at the limit $\mu \rightarrow 0$ the  
retarded perturbations $h^{(1)}_{\mu\nu}$  
are identical to the retarded first-order 
perturbations that are induced by a unit-mass point particle tracing  
the same worldline $z(\tau)$. 

Before tackling Eqs. (\ref{newfst},\ref{newsnd})
we must provide additional information about $z(\tau)$.
Recall that as $\mu\rightarrow 0$ the worldtube $S$ collapses to  
the worldline $z(\tau)$.
Roughly speaking one may choose $S$ to follow 
the motion of the black-hole keeping it ''centered'' at all times 
with respect to $S$ in some well defined manner \footnote{Within the internal zone 
the full spacetime metric is approximated by the metric of 
a perturbed Schwarzschild black hole. 
Here fixing the center of the black hole amounts to fixing 
the internal-zone dipole perturbations (These dipole perturbations are purely gauge and one 
can therefore set them to zero.) \cite{MST}.}. 
This point of view allows one 
to identify $z(\tau)$ as a representative ''world-line'' 
of the black-hole in the background spacetime. 
At the leading order of approximation 
the representative 
worldline of the black-hole was found to be a geodesic in the background spacetime
 (see e.g., \cite{DEATH,MST}). 
Alternatively one may let the black-hole
drift with respect to the center of $S$. 
In this article we find that this alternative point of view more suitable 
for our purposes of studying the second-order perturbations. The main reason 
is that it allows us to choose $z(\tau)$ to be exactly a geodesic in 
the background spacetime which we denote with $z_G(\tau)$.
Setting $z(\tau)=z_G(\tau)$ guarantees that Eq. (\ref{newfst})
has an exact retarded solution [given by Eq. (\ref{h1ret}) below]. 
Notice that if we had chosen a point of view where
$z(\tau)$ represents the (generically accelerated) motion of the black-hole 
we would have found that 
 Eq. (\ref{newfst}) does not have any exact solution.
This difficulty originates form the fact 
that application of the divergence operator
to left hand side of Eq. (\ref{newfst}) gives
 $\nabla^\nu R_{\mu\nu}^{(L)}[h^{(1)}]\equiv 0$, 
which restricts the possible sources allowed on the right hand side.
As we already mentioned, matching with the internal-zone solution implies  
that one may replace the right hand side of Eq. (\ref{newfst}) with a point particle 
(delta-function) source at $z(\tau)$. Here we find that the 
above mentioned restriction on the source implies that 
 $z(\tau)$ must be a geodesic worldline. 
However, the motion of the black-hole in the background spacetime  
is generically an accelerated motion. This acceleration originates 
from the gravitational self-force acting on the small black-hole. 
Therefore, had we chosen $z(\tau)$ to represent the (accelerated) motion 
of the black-hole we would have found
that Eq. (\ref{newfst}) does not have any exact solution.
Our point of view different, since we set $z(\tau)=z_G(\tau)$
and therefore Eq. (\ref{newfst}) has a well defined exact solution. 
The black-hole acceleration which 
will gradually shift the black-hole from the center of the worldtube $S$ 
will show up as a term in 
the boundary conditions for the external-zone solution as
$x\rightarrow z_G(\tau)$.
Notice that since the leading order 
acceleration scales like $\mu$ the difference in the boundary
conditions induced by this leading-order acceleration
will affect only  $h^{(2)}_{\mu\nu}$, thus placing the effect of the 
first-order self-force at the boundary conditions for the second-order equation
 (\ref {newsnd}).

With the above choice of worldline 
the equations for $h^{(1)}_{\mu\nu}$ and $h^{(2)}_{\mu\nu}$ now
read 
\begin{eqnarray}\label{finalfst}
R_{\mu\nu}^{(L)}[h^{(1)}]=0&\,\,\,\,\,,x \not\in z_G(\tau) \,,\\ \label{finalsnd}
R_{\mu\nu}^{(L)}[h^{(2)}]=-R_{\mu\nu}^{(Q)}[h^{(1)}]&\,\,\,\,\,,x \not\in z_G(\tau)
\,.
\end{eqnarray}

\section{First-order metric perturbations}\label{fop}

First, we consider the construction of the first-order metric perturbations 
$h_{\mu\nu}^{(1)}$, which satisfy Eq. (\ref{finalfst}). 
As was previously mentioned, 
 at the limit $\mu \rightarrow 0$ the perturbations $h_{\mu\nu}^{(1)}$ are identical 
to the first-order metric perturbations induced by a unit-mass point-particle  
which traces a geodesic $z_G(\tau)$ on the vacuum background metric.
We impose the Lorenz-gauge conditions on these first-order
 perturbations, reading
\[
\bar{h}^{(1)\mu\nu}_{\ \ \ \ \ ;\nu}=0\, ,
\]
where overbar denotes the trace-reversal operator 
defined by
 $\bar{h}^{(1)}_{\mu\nu}\equiv h^{(1)}
_{\mu\nu} - 1/2g_{\mu\nu}h_\alpha^{(1)\alpha}$.
In this gauge the first-order perturbations $h_{\mu\nu}^{(1)}$
read (see e.g. \cite{PR})
\begin{equation}\label{h1ret}
\bar{h}^{(1)}_{\mu\nu}(x)
=4\int_{-\infty}^{\infty}
 G_{\mu\nu\alpha\beta}^{ret}[x|z_G(\tau)]u^\alpha(\tau)u^\beta(\tau) d\tau\,.
\end{equation} 
Here $u^\alpha\equiv\frac{dz_G^\alpha}{d\tau}$, and 
 $G^{ret}_{\mu\nu\alpha\beta}[x|z_G(\tau)]$
 is the gravitational retarded Green's function which is a bi-tensor, where
 the indices $\alpha,\beta$ refer to $z(\tau)$, and the indices $\mu,\nu$
  refer to $x$. This Green's function satisfies 
\[
\Box G^{\mu\nu}_{\ \ \alpha'\beta'}[x|x']+2R_{\eta\ \rho}^{\ \mu\ \nu}(x)
G^{\eta\rho}_{\ \ \alpha'\beta'}[x|x']=-4\pi{\bar{g}}^{(\mu}_{\ \alpha'}(x,x')
{\bar{g}}^{\nu)}_{\beta'}(x,x'){[-g]}^{-1/2}\delta^4(x-x')\,.
\]
Here $\Box\equiv g^{\rho\sigma}\nabla_{\rho}\nabla_{\sigma}$ is a differential
     operator at $x$,
 ${\bar{g}}^{\mu}_{\alpha'}(x,x')$ denotes the bi-vector of a geodesic 
parallel transport with respect to the background metric (for the 
properties of this bi-vector see e.g. \cite{DB,PR}),
$R_{\eta\mu\rho\nu}$ denotes Riemann tensor of the background geometry
with the sign convention of reference \cite{MTW}, $g$ denotes the determinant 
of the background metric, and $\delta^4(x-x')$ denotes the four-dimensional 
(coordinate) Dirac delta-function. Throughout this article we use 
the signature $(-,+,+,+)$ and geometric units units $G=c=1$
.   
\section{Second-order metric perturbations}\label{sop}

We now focus our attention to the construction of the second-order perturbations 
$h_{\mu\nu}^{(2)}$ which satisfy Eq. (\ref{finalsnd}).
Here it will be useful to apply the trace reverse operator to Eq. (\ref{finalsnd})
which gives $\bar{R}_{\mu\nu}^{(L)}[h^{(2)}]=-\bar{R}_{\mu\nu}^{(Q)}[h^{(1)}]$.
To simplify the notation we rewrite this equation as 
\begin{equation}\label{h2simple}
D_{\mu\nu}[\bar{h}^{(2)}]=S_{\mu\nu}[\bar{h}^{(1)}]\ \ \ ,x\not\in z_G(\tau) .
\end{equation}
Here we substituted 
$h^{(2)}_{\mu\nu}=\bar{h}^{(2)}_{\mu\nu}-(1/2)g_{\mu\nu}\bar{h}^{(2)\alpha}_{\ \ \ \ \alpha}$
 into $\bar{R}_{\mu\nu}^{(L)}[h^{(2)}]$ and denoted the 
resultant expression with $D_{\mu\nu}[\bar{h}^{(2)}]$ 
(notice that $\bar{R}_{\mu\nu}^{(L)}[h^{(2)}]\ne R_{\mu\nu}^{(L)}[\bar{h}^{(2)}]$).
Similarly the source term $S_{\mu\nu}[\bar{h}^{(1)}]$ is defined 
by $S_{\mu\nu}[\bar{h}^{(1)}]\equiv  -\bar{R}_{\mu\nu}^{(Q)}[h^{(1)}]$; 
and for abbreviation we shall often simply write $S_{\mu\nu}$.
The explicit form of these terms is provided in appendix A. For the 
moment we do not impose any second-order gauge conditions and Eq. (\ref{h2simple})
 is in a general second-order gauge. 

Before constructing a solution to 
Eq. (\ref{h2simple}) let us first study the singular properties of $S_{\mu\nu}$ near
$z_G(\tau)$. In what follows we shall expand $\bar{h}^{(1)}_{\mu\nu}$
 and $S_{\mu\nu}$ in the vicinity of $z_G(\tau)$. Throughout this paper
such tensor expansions are considered
on a family of hypersurfaces $\tau=const$
that are generated by geodesics which are normal to 
worldline. Each point $x$ on such a hypersurface  
is associated with the same point on the worldline $z_G(\tau_x)$. 
On each of these hypersurfaces the expansions 
are valid only in a local neighborhood of $z_G(\tau_x)$ 
excluding a sphere of arbitrarily small volume which surrounds $z_G(\tau_x)$. 
Here it should be noticed that since the 
dynamical equations (\ref{finalfst},\ref{finalsnd}) are not valid on the 
worldline we do not have to keep track of the singularities on the worldline, for example 
 distributions of the form $\delta^3(x-z_G)$ which may arise 
in the tensor expansions below due to application of a Laplacian 
operator on terms of the form $1/r$  
may be completely discarded. Throughout this paper 
(unless we explicitly indicate otherwise) 
we shall represent the expansions of tensor fields 
using Fermi normal coordinates based on $z_G(\tau)$.
These expansions take a particularly simple form in these coordinates, 
and often many of the leading terms are found to be identical 
to the corresponding terms in an expansion over a flat background spacetime using 
Lorentz coordinates. We shall use the symbol $\stackrel{*}=$ to denote equality 
in a particular coordinate system.
Note that the covariant nature of 
Eqs. (\ref{finalfst},\ref{finalsnd}) implies that 
working in a particular (background) coordinate system does not reduce the generality 
of our analysis, since once a solution is constructed in one 
particular coordinate system 
it can be transformed to any other coordinate system by 
a coordinate transformation.
Moreover, our results which provide a prescription for constructing the 
second-order gravitational perturbations are stated in a covariant manner, 
and therefore can be implemented in any coordinate system.

Using  Eq. (\ref{h1ret}) we expand 
$\bar{h}^{(1)}_{\mu\nu}$ in the vicinity of $z_G(\tau)$, which gives 
\begin{equation}\label{h1dec}
\bar{h}^{(1)}_{\mu\nu}(x)\stackrel{*}=4u_\mu u_\nu r^{-1}
+O(r^0)\,.
\end{equation}
In the external-zone $r$ denotes the invariant spatial distance along a geodesic connecting  
$z(\tau_x)$ and $x$, $r=\sqrt{\delta_{ab}x^a x^b}$ where $x^a$ are the 
spatial Fermi coordinates; $u^\mu\stackrel{*}=\delta^\mu_0$ is 
a vector field which coincide with four-velocity on the worldline. 
By substituting Eq. (\ref{h1dec}) into $S_{\mu\nu}$ we 
obtain the following expansion
\begin{equation}\label{sourcexpan}
S_{\mu\nu}(x)\stackrel{*}=\left[4u_{\mu}u_{\nu}+7\eta_{\mu\nu}-14\Omega_{\mu}\Omega_{\nu}
 \right]r^{-4}+O(r^{-3})\,.
\end{equation}
Here $\eta_{\mu\nu}$ denotes Minkowski metric, and we defined $\Omega^a\stackrel{*}=x^a/r$,
 $\Omega^0\stackrel{*}=0$, and $\Omega_\mu\stackrel{*}=g_{\mu\nu}\Omega^\nu$.
Naively one may try to construct the standard retarded solution to 
Eq. (\ref{h2simple}), say by  
imposing Lorenz gauge conditions on $h^{(2)}_{\mu\nu}$ and then formally integrating  
$S_{\mu\nu}$ with the retarded Green's function which gives
\begin{equation}\label{naive}
\bar{h}^{(2)}_{\mu\nu}(x)
=\frac{1}{2\pi}\int_{-\infty}^{\infty}
 G_{\mu\nu\alpha'\beta'}^{ret}[x|x']S^{\alpha'\beta'}(x')\sqrt{-{g(x')}} d^4x' \,.
\end{equation} 
Here there is a problem, examining 
Eq. (\ref{sourcexpan}) reveals that $S_{\mu\nu}$ diverges like 
$r^{-4}$ in the vicinity of the worldline $z_G(\tau)$. 
Recalling  that $\sqrt{{-g(x')}} d^4x'$ scales like
$r^2$, we find that the integral in Eq. (\ref{naive}) diverges at 
 every point in spacetime.
Furthermore, the next order term in Eq. (\ref{sourcexpan}) that
diverges like $r^{-3}$ also gives rise to a divergent integral.

We will now develop a method to obtain well defined solutions 
for Eq. (\ref{h2simple}). This method
is based on consecutive steps, where in each 
step we reduce the degree of singularity of the field equation at hand.
Eventually we end up with a field equation for a certain
 residual potential which has a source term
which diverges like $r^{-2}$, this equation has well defined
retarded solutions. At this point we shall construct a retarded solution to this 
equation and discuss the matching to the internal-zone solution.

We should mention here another 
difficulty in calculating the integral in Eq. (\ref{naive}). 
Asymptotically $h^{(1)}$ 
has a form of a gravitational wave. 
Therefore, the leading asymptotic behavior of 
$\nabla h^{(1)}$ is $O(R^{-1})$, where $R$ is 
 the area coordinate - in this paragraph, for simplicity,
 we specialize to a background metric of a Schwarzschild black hole.
This implies that 
 for an infinitely long 
world line the source term $S_{\alpha\beta}$ decays asymptotically 
as $R^{-2}$. This term has a static  $O(R^{-2})$ part
which does not vanish after time averaging.
An attempt to calculate the integral (\ref{naive}) over this static $O(R^{-2})$
part produces a divergent integral (even if we resolve the difficulty with 
the singularity near the world line).
The regularization of this divergency lies outside 
the scope of this article (Ori has recently suggested a resolution to 
this problem \cite{Ori}). 
In what follows we shall assume that such a regularization at 
infinity has been carried out.

\subsection{$r^{-4}$ singularity}

First we tackle the strongest ($r^{-4}$) singularity in the source term of Eq. (\ref{h2simple}).
For this purpose let us
 decompose $\bar{h}^{(2)}_{\mu\nu}$ into two tensor potentials, reading 
\begin{equation}
\bar{h}^{(2)}_{\mu\nu}=\bar{\psi}_{\mu\nu}+\delta \bar{h}^{(2)}_{\mu\nu}\,,
\end{equation}
where $\bar{\psi}_{\mu\nu}$ satisfies
\begin{equation}\label{psieq}
D_{\mu\nu}[\bar{\psi}]\stackrel{*}=\left[4u_{\mu}u_{\nu}+7\eta_{\mu\nu}-14\Omega_{\mu}\Omega_{\nu}
 \right]r^{-4}+O(r^{-3})\,.
\end{equation}
Notice that the $r^{-4}$ singular term in Eq. (\ref{psieq}) is the same as
the $r^{-4}$ singular term in the expansion of $S_{\mu\nu}$
 [see Eq. (\ref{sourcexpan})],
 but the lower order terms of equations (\ref{h2simple}) and (\ref{psieq}) 
are in general different. In fact, we do not
impose any restrictions on the lower order terms in Eq. (\ref{psieq}).
Suppose that we construct a solution to Eq. (\ref{psieq}), then by subtracting
$D_{\mu\nu}[\bar{\psi}]$ from both sides of Eq. (\ref{h2simple}) we obtain the following 
equation for $\delta \bar{h}_{\mu\nu}^{(2)}$
\begin{equation}\label{deltah2eq}
D_{\mu\nu}[\delta \bar{h}^{(2)}]=S_{\mu\nu}-D_{\mu\nu}[\bar{\psi}]\ \ \  ,x\not\in z_G(\tau)\,.
\end{equation}
By construction the source term in
this equation diverges only like $r^{-3}$, while the original field 
equation (\ref{h2simple}) has a source term which diverges like $r^{-4}$. In this 
sense Eq. (\ref{deltah2eq}) is simpler then Eq. (\ref{h2simple}).

We now face the problem of solving Eq. (\ref{psieq}). 
To construct a particular solution 
we use a linear combination of 
terms which are quadratic in $\bar{h}^{(1)}_{\mu\nu}$.
Since $\bar{h}^{(1)}_{\mu\nu}$ diverges like $r^{-1}$ we find 
that by applying the differential operator $D_{\mu\nu}$ to terms
which are quadratic in $\bar{h}^{(1)}_{\mu\nu}$, we can obtain 
terms which diverge like $r^{-4}$. First we construct four independent
quadratic tensor fields reading
\begin{eqnarray}\label{quadratic}
&&\varphi^{A}_{\mu\nu}=\bar{h}^{(1)\rho}_{\ \ \ \ \mu}
\bar{h}^{(1)}_{\rho\nu}\ \ \ \ \ \ \ \ \ \ \  ,\ \ 
\varphi^{B}_{\mu\nu}=\bar{h}^{(1)\rho}_{\ \ \ \ \rho}
\bar{h}^{(1)\ \ \ \ }_{\mu\nu}\\ \nonumber
&&\varphi^{C}_{\mu\nu}=\bar{h}^{(1)\eta\rho}
\bar{h}^{(1)}_{\ \ \ \eta\rho}g_{\mu\nu}\ \ ,\ \ 
\varphi^{D}_{\mu\nu}=\left(\bar{h}^{(1)\rho}_{\ \ \ \ \rho}\right)^2
g_{\mu\nu}\,.
\end{eqnarray}
These terms can be combined to form a solution to Eq. (\ref{psieq}) which reads
\begin{equation}\label{psiexplicit}
\bar{\psi}_{\mu\nu}=\frac{1}{64}\left[2(c_A\varphi^{A}_{\mu\nu}+
c_B\varphi^{B}_{\mu\nu})-7(c_C\varphi^{C}_{\mu\nu}+
c_D\varphi^{D}_{\mu\nu})\right]\,.
\end{equation}
Here  the constants $c_A,c_B,c_C,c_D$ must satisfy
\begin{equation}\label{cacbcccd}
c_A+c_B=1\ \ \ ,\ \ \ c_C+c_D=1\, ,
\end{equation}
but are otherwise arbitrary. 
One may directly substitute Eq. (\ref{psiexplicit}) 
into Eq. (\ref{psieq}) and verify that 
$\bar{\psi}_{\mu\nu}$ satisfies this equation.

Eq. (\ref{psiexplicit}) can be derived as follows.
First notice that the coefficient in front of the $r^{-4}$ term 
in both Eq. (\ref{sourcexpan}) and Eq. (\ref{psieq}) 
does not depend on the curvature of the background spacetime.
In fact the form of this expression would not change
 if we would replace the curved background spacetime
with a flat spacetime. 
Therefore in deriving $\bar{\psi}_{\mu\nu}$ we may consider a simple 
case of flat background spacetime.
In this case, 
the exact nonlinear solution to Einstein's field equations 
in our problem is simply the Schwarzschild solution.  
Far from the black-hole this Schwarzschild solution 
may be approximated by an expansion which schematically reads
\begin{equation}\label{schexp}
g_{\mu\nu}^{Sch}=\eta_{\mu\nu}+\mu H^{(1)}_{\mu\nu}+\mu^2 H^{(2)}_{\mu\nu}+ O(\mu^3)\,.
\end{equation}
In the appropriate coordinates the second-order term $\bar{H}^{(2)}_{\mu\nu}$ 
satisfies Eqs. (\ref{h2simple},\ref{psieq}) 
in the flat background case.
To be consistent with our first-order (Lorenz) gauge conditions we have to make sure that 
the term $H^{(1)}_{\mu\nu}$ satisfies the Lorenz gauge conditions 
for a flat spacetime metric. As we will 
immediately show this condition is satisfied if we express the Schwarzschild 
solution in isotropic Cartesian coordinates. 
The Schwarzschild metric in the isotropic Cartesian coordinates takes the form of
\begin{equation}\label{isotropic}
ds^2=-\left(\frac{2\tilde{r}-\mu}{2\tilde{r}+\mu}\right)^2dt^2+
\left(1+\frac{\mu}{2\tilde{r}} \right)^4(dx^2+dy^2+dz^2)\,.
\end{equation}
Here $\tilde{r}^2=x^2+y^2+z^2$. Expanding this metric in powers of $\mu/\tilde{r}$ gives
the following expressions for the trace-reversed 
first-order and second-order perturbations 
\begin{equation}\label{bigh}
\bar{H}^{(1)}_{\mu\nu}\stackrel{*}=\frac{4}{\tilde{r}}\tilde{u}_\mu \tilde{u}_\nu
 \ \ \ \ , \ \ \ \
 \bar{H}^{(2)}_{\mu\nu}\stackrel{*}=-\frac{1}{4\tilde{r}^2}
\left[2\tilde{u}_\mu \tilde{u}_\nu+7\eta_{\mu\nu}\right]\,.  
\end{equation}
Here $\tilde{u}^{\mu}\stackrel{*}=\delta^\mu_0$ is a vector field.  
Notice that the Lorenz gauge conditions 
$\bar{H}^{(1)\mu\nu}_{\ \ \ \ \ ,\nu}\stackrel{*}=0$ 
are satisfied. We may replace $\bar{h}^{(1)}_{\mu\nu}$ with $\bar{H}^{(1)}_{\mu\nu}$  
in the quadratic terms (\ref{quadratic}) and employ  Eqs. (\ref{bigh}) 
to express $\bar{H}^{(2)}_{\mu\nu}$ as a linear combination of 
these quadratic terms. The coefficients of this linear combination are 
the desired coefficients in Eqs. (\ref{psiexplicit},\ref{cacbcccd}).
Notice that for the case of a flat background space-time $\bar{\psi}_{\mu\nu}$ is an 
{\em exact solution} to Eq. (\ref{h2simple}).
\subsection{$r^{-3}$ singularity}

We now consider the construction of $\delta \bar{h}_{\mu\nu}^{(2)}$
which satisfies  Eq. (\ref{deltah2eq}). This equation 
has a source term which diverges like $r^{-3}$, and therefore
its standard retarded solution diverges. 
Let us examine the terms that give rise to this $r^{-3}$ singularity. 
It is convenient to express
the source term of  Eq. (\ref{deltah2eq})
schematically (and without indices) as 
\begin{equation}\label{deltah2source}
S-D[\bar{\psi}]=\nabla\bar{h}^{(1)}\nabla\bar{h}^{(1)}\ 
\& \ \bar{h}^{(1)}\nabla\nabla\bar{h}^{(1)}\,.
\end{equation}
Notice that here only sum over all the  
terms diverges as $r^{-3}$ (Since the individual terms which diverge as  
 $r^{-4}$ cancel each other.).
Using a decomposition devised by Detweiler and Whiting \cite{DW}
we decompose $\bar{h}^{(1)}_{\mu\nu}$ as follows\footnote{Here 
we define  $\bar{h}^{(1)R}_{\mu\nu}$ and $\bar{h}^{(1)S}_{\mu\nu}$ 
to be independent of  $\mu$, as opposed to 
\cite{DW} .}
\begin{equation}\label{h1decomp}
\bar{h}^{(1)}_{\mu\nu}=\bar{h}^{(1)S}_{\mu\nu}+\bar{h}^{(1)R}_{\mu\nu}\,.
\end{equation}
Here $\bar{h}^{(1)S}_{\mu\nu}$ is a certain singular potential which diverges 
as $r^{-1}$ as $r\rightarrow 0$, and $\bar{h}^{(1)R}_{\mu\nu}$ is 
a certain regular potential which satisfies the following homogeneous wave equation
\begin{equation}\label{h1req}
\Box \bar{h}^{(1)R}_{\mu\nu}+2R^{\eta\ \rho}_{\ \mu\ \nu}\bar{h}^{(1)R}_{\eta\rho}=0 \,.
\end{equation}
Decomposition (\ref{h1decomp}) is particularly useful 
for expressing the first-order gravitational 
self-force, since the general expression 
of this self-force is completely determined from $\bar{h}^{(1)R}_{\mu\nu}$ [see \cite{DW} 
and also Eq. (\ref{selfaccel}) below].
Generically $\bar{h}^{(1)R}_{\mu\nu}$ is a smooth field of 
$O(r^0)$ on $z_G(\tau)$.
Expanding  $\bar{h}^{(1)S}_{\mu\nu}$ and its covariant derivatives 
in the vicinity of the worldline $z_G(\tau)$ gives (A method of constructing these 
expressions is described in detail in \cite{PR})
\begin{eqnarray}\label{h1s}
&&\bar{h}^{(1)S}_{\mu\nu}\stackrel{*}=\frac{4}{r}u_\mu u_\nu+O(r^1)
\\\label{gradh1s}
&&\nabla_\rho\bar{h}^{(1)S}_{\mu\nu}\stackrel{*}=-\frac{4}{r^2}u_\mu u_\nu\Omega_\rho+O(r^0)
\\\label{2gradh1s}
&&\nabla_\eta\nabla_\rho\bar{h}^{(1)S}_{\mu\nu}\stackrel{*}=
\frac{4u_{\mu}u_{\nu}}{r^3}\left[3\Omega_{\eta}\Omega_{\rho}-u_{\eta}u_{\rho}-\eta_{\eta\rho}
\right]+O(r^{-1})\,.
\end{eqnarray}
Notice that the orders $r^0$, $r^{-1}$ and $r^{-2}$ are missing from the expansions
of $\bar{h}^{(1)S}_{\mu\nu}$, $\nabla_\rho\bar{h}^{(1)S}_{\mu\nu}$ 
 and $\nabla_\eta\nabla_\rho\bar{h}^{(1)S}_{\mu\nu}$,
respectively. The absence of these terms can be 
traced to the vanishing acceleration of $z_G(\tau)$.
We now substitute Eq. (\ref{h1decomp}) into Eq. (\ref{deltah2source}) and 
examine the various terms that give rise to the problematic $r^{-3}$ 
singularity in the source term $S-D[\bar{\psi}]$. 
Expansions (\ref{h1s},\ref{gradh1s},\ref{2gradh1s}) 
imply that the only combination that produces this $r^{-3}$ singularity is 
of the form $\bar{h}^{(1)R}\nabla\nabla\bar{h}^{(1)S}$.

To eliminate this problematic $r^{-3}$ singularity we utilize gauge freedom and
employ a first-order (regular) gauge transformation $x^\nu \rightarrow x^\nu-\mu \xi^\nu$, 
which gives
\begin{equation}\label{newhr}
h^{(1)}_{\mu\nu}\rightarrow h^{(1)S}_{\mu\nu}+h^{(1)R(new)}_{\mu\nu}\ \ , \ \ 
h^{(1)R(new)}_{\mu\nu}\equiv h^{(1)R}_{\mu\nu}+\xi_{\mu;\nu}+\xi_{\nu;\mu}\,.
\end{equation}
Here $\xi_{\nu}$ does not depend on $\mu$.
Notice that we have included the entire gauge transformation 
in the definition of the new regular potential $h^{(1)R(new)}_{\mu\nu}$. 
This is a natural identification since the gravitational 
self-force in the new-gauge is obtained by replacing $h^{(1)R}_{\mu\nu}$
with $h^{(1)R(new)}_{\mu\nu}$ in the expression of the self-force 
\cite{sfgauge}. 
We now impose the following gauge conditions 
\begin{equation}\label{1gaugecon}
[h^{(1)R(new)}_{\mu\nu}]_{z_G(\tau)}\equiv
\biglb[h^{(1)R}_{\mu\nu}+\xi_{\mu;\nu}+\xi_{\nu;\mu}\bigrb]_{z_G(\tau)}=0\,.
\end{equation}
Expanding $h^{(1)R(new)}_{\mu\nu}$ in the vicinity of the worldline $z_G(\tau)$ gives 
\[
h^{(1)R(new)}_{\mu\nu}=O(r)\,.
\]
Most beneficially in this new gauge the previously mentioned 
problematic terms $\bar{h}^{(1)R(new)}\nabla\nabla\bar{h}^{(1)S}$ in the source term 
$S-D[\bar{\psi}]$ diverge only like $r^{-2}$. This property will allow us 
to construct well defined retarded solution to Eq. (\ref{deltah2eq}).
Notice that we invoked a regular gauge transformation in the sense that it
 did not change the singular properties of 
$\bar{h}^{(1)}_{\mu\nu}$ near the worldline, meaning that 
Eq. (\ref{h1dec}) is unchanged by 
this gauge transformation. Therefore, the coefficient in front of the $r^{-4}$ term in 
Eqs. (\ref{h2simple},\ref{psieq}) is not affected by the gauge transformation. 
This implies that even though the numerical values of 
$\bar{\psi}_{\mu\nu}$ are changed by the gauge transformation, the general form 
of $\bar{\psi}_{\mu\nu}$ given by Eqs. (\ref{psiexplicit},\ref{cacbcccd}) 
is invariant to any such regular gauge transformations [e.g., a transformation 
satisfying Eq. (\ref{1gaugecon})].

\subsection{Construction of the first-order gauge}\label{worldline}

As was previously mentioned the first-order gravitational self-force 
must be accounted for when imposing boundary conditions as $x\rightarrow z_G(\tau)$ 
for Eq. (\ref{finalsnd}).
To simplify the calculation of these boundary conditions we once 
more use the gauge freedom.
The first-order gravitational self-force is a gauge dependent 
quantity \cite{BOGAUGE}, and in fact one can always choose a convenient first-order 
gauge in which the first-order gravitational self-force vanishes (see below). 
In this gauge the geodesic worldline $z_G(\tau)$ represents 
the black-hole's worldline accurately up to errors of order $\mu^2$. In this case the 
contributions to the boundary conditions of Eq. (\ref{finalsnd}) that originate from the 
black-hole's acceleration due to the first-order self-force simply vanish.

To spell out the desired gauge conditions let us 
examine the expression for the $O(\mu)$ acceleration which is 
induced by the first-order self-force \cite{DW}
\begin{equation}\label{selfaccel}
a^{\mu}=-\mu(g^{\mu\nu}+u^{\mu}u^{\nu})u^{\rho}u^{\eta}
(\nabla_\rho h^{(1)R}_{\eta\nu}-\frac{1}{2}\nabla_{\nu}h^{(1)R}_{\rho\eta})\,.
\end{equation}
Here all quantities are evaluated on the worldline.
Originally  this expression was derived for $h^{(1)R}_{\mu\nu}$ which 
satisfies the Lorenz gauge conditions. 
But following the analysis in \cite{BOGAUGE} we find that this expression 
is also valid in any new gauge provided that the gauge transformation from Lorenz gauge 
to this new gauge is sufficiently smooth. 
To obtain a gauge with a vanishing first-order gravitational self-force we should
require that $a^{\mu}=0$. This requirement conforms with many gauge choices. For 
example it is satisfied if all the first covariant derivatives of 
the regular field in the new gauge vanish. 
Putting this gauge condition together 
with our previous gauge condition (\ref{1gaugecon})
yields 
\begin{eqnarray}\label{1fermigauge}
&&\left[h^{(1)R(Fermi)}_{\mu\nu}\right]_{z_G(\tau)}\equiv\biglb[h^{(1)R}_{\mu\nu}+\xi_{\mu;\nu}+\xi_{\nu;\mu}\bigrb]_{z_G(\tau)}=0\,,
\\\label{2fermigauge}
&& \left[\nabla_\rho h^{(1)R(Fermi)}_{\mu\nu}\right]_{z_G(\tau)}\equiv\Biglb[\nabla_\rho\biglb(
h^{(1)R}_{\mu\nu}+\xi_{\mu;\nu}+\xi_{\nu;\mu}\bigrb)\Bigrb]_{z_G(\tau)}=0\,.
\end{eqnarray}
We shall refer to this new gauge as Fermi gauge.

To construct Fermi gauge consider 
contracting Eq. (\ref{2fermigauge}) with $u^\rho$. The resultant equation 
states that $h^{(1)R(Fermi)}_{\mu\nu}$ is constant along $z_G(\tau)$, 
which is consistent with Eq. (\ref{1fermigauge}). 
But more importantly it implies that once Eq. (\ref{1fermigauge}) is satisfied at an 
initial point $z_G(\tau_0)$ Eq. (\ref{2fermigauge}) will guarantee its validity 
everywhere along $z_G(\tau)$. We now choose an arbitrary gauge vector $\xi_{(0)\mu}$ 
at some initial point $z_G(\tau_0)$, and construct its first 
covariant derivatives at this point 
such that Eq. (\ref{1fermigauge}) is satisfied.  
For example we may choose 
\begin{equation}\label{initcon}
\xi_{(0)\mu;\nu}=\xi_{(0)\nu;\mu}=-\frac{1}{2} h^{(1)R}_{\mu\nu}(\tau_0)\,.
\end{equation}
To transport $\xi_{\mu}$  and  $\xi_{\mu;\nu}$ along $z_G(\tau)$ we derive 
transport equations as follows. We treat Eq. (\ref{2fermigauge}) 
and the commutation relation 
$2\xi_{\mu;[\nu\alpha]}=R^{\epsilon}_{\ \mu\nu\alpha}\xi_{\epsilon}$ 
as a set of algebraic equations for $\xi_{\alpha;\beta\gamma}$ and use
the identities of Riemann tensor to obtain the following relation 
\begin{equation}\label{gradgradxi}
\xi_{\nu;\mu\alpha}=R^{\epsilon}_{\ \alpha\mu\nu}\xi_{\epsilon}
-\frac{1}{2}\biglb(h^{(1)R}_{\mu\nu;\alpha}+h^{(1)R}_{\nu\alpha;\mu}-
h^{(1)R}_{\mu\alpha;\nu}\bigrb)\,.
\end{equation}
Here all quantities are evaluated on the worldline. One may  
substitute Eq. (\ref{gradgradxi}) into Eq. (\ref{2fermigauge}), and 
into the above mentioned  commutation relation; and thereby verify that these
two equations are identically satisfied by Eq. (\ref{gradgradxi}). 
We now construct a second-order transport equations for $\xi_{\mu}(\tau)$ by 
contracting Eq. (\ref{gradgradxi}) with $u^\alpha u^\mu$ which 
gives
\begin{equation}\label{transxi}
\frac{D^2}{D\tau^2}\xi_\nu=R^{\epsilon}_{\ \alpha\mu\nu}\xi_{\epsilon}u^\alpha u^\mu-
\frac{1}{2}u^\alpha u^\mu\biglb(h^{(1)R}_{\mu\nu;\alpha}+h^{(1)R}_{\nu\alpha;\mu}-
h^{(1)R}_{\mu\alpha;\nu}\bigrb)\,.
\end{equation}
Solving this equation with the above mentioned initial conditions 
provides us with the gauge vector $\xi_{\mu}(\tau)$  along the worldline. 
Similarly we can construct a first-order transport equation for $\xi_{\mu;\nu}$ 
by contracting Eq. (\ref{gradgradxi}) with $u^\alpha$. 
Using the solution $\xi_{\mu}(\tau)$ of Eq. (\ref{transxi}) together with  
the initial conditions (\ref{initcon}) this first-order transport 
equation can be integrated to give $\xi_{\mu;\nu}(\tau)$. $\xi_{\mu;\nu\alpha}(\tau)$
is then obtained by substituting $\xi_{\mu}(\tau)$ into Eq. (\ref{gradgradxi}).  
Once $\xi_\mu(\tau)$, $\xi_{\mu;\nu}(\tau)$, and $\xi_{\mu;\nu\alpha}(\tau)$ are obtained, 
one can use these quantities to construct a local expansion of 
the gauge vector field $\xi^\mu(x)$ in a local neighborhood of the worldline.
Notice that Eq. (\ref{gradgradxi}) implies that Fermi gauge satisfies the 
Lorenz gauge conditions along the worldline.

The above construction only 
provides leading terms in an expansion of  $\xi_\mu$ in a local neighborhood of the worldline.
One may continue $\xi_\mu$ globally to the entire spacetime. 
Since gauge freedom is associated with non-physical degrees of 
freedom we are allowed to introduce a gauge continuation which 
depends on arbitrary parameters.
Nevertheless, it is sometimes helpful to work in a gauge which 
is manifestly causal, such a gauge may be constructed by continuing $\xi_\mu$
 along future null cones based on $z_G(\tau)$ \cite{nullcont}. 
In the analysis below we assume that such a global gauge 
continuation has been performed and that the first-order gauge is fixed globally.

\subsection{Particular solution $\delta \bar{h}^{(2)}$}

We shall now construct a particular retarded solution to Eq. (\ref{deltah2eq}). 
Here it is useful to remove the restriction $x\not\in z_G(\tau)$,
and continue the source of Eq. (\ref{deltah2eq}) to the world line.
Clearly a particular solution to  Eq. (\ref{deltah2eq}) with the world line 
included also satisfies the original 
equation (i.e. with the worldline excluded). 
However, not every continuation of the source of Eq. (\ref{deltah2eq}) to the world-line
produces an equation which is self-consistent. 
This is easily demonstrated by taking 
the covariant divergence of both sides of Eq. (\ref{deltah2eq}). 
In appendix B we show that
$\nabla^{\mu}D_{\mu\nu}[\delta \bar{h}^{(2)}]$ vanishes identically, 
and furthermore the covariant divergence of the source term of Eq. (\ref{deltah2eq}) in  
$x\not\in z_G(\tau)$ vanishes as well. 
These facts constrain the permitted continuations of the source 
of Eq. (\ref{deltah2eq}) to the worldline. 
If one naively chooses a continuation to the worldline
which has a non-vanishing covariant divergence, the resulted 
equation will not be self consistent.
Here we 
choose the simplest possible continuation by requiring 
that no additional singularities are introduced on the worldline. Meaning 
that the singularities of the continued source term  
on the worldline are completely specified by its expansion
in $x\not\in z_G(\tau)$, and no additional singularities (e.g. delta functions) 
are introduced on the worldline.
Eq. (\ref{deltah2eq}) now takes the form
\begin{equation}\label{fermideltah2eq}
D_{\mu\nu}[\delta \bar{h}^{(2)}]=\delta S^F_{\mu\nu}\,.
\end{equation}
Here $\delta S^F_{\mu\nu}\equiv S^F_{\mu\nu}-D_{\mu\nu}[\bar{\psi}^F]$, where
 the superscript $F$ indicates that source terms are evaluated in Fermi gauge.
Below we show that  Eq. (\ref{fermideltah2eq}) is self-consistent 
by constructing a solution to this equation.

Recall that we have fixed the first-order gauge, but we still have the freedom to invoke
a purely second-order gauge transformation of the form 
\[
x^\mu\rightarrow x^{\mu}-\mu^2 \xi^{\mu}_{(2)}\,.
\]
Here the gauge vector $\xi^{\mu}_{(2)}$ is independent of the mass $\mu$.
Similar to the first-order case, 
one may choose the gauge vector $\xi^{\mu}_{(2)}$
such that $\delta\bar{h}^{(2)}_{\mu\nu}$  satisfies 
the Lorenz gauge conditions, reading 
\begin{equation}\label{lg}
\delta\bar{h}^{(2)\mu\nu;}_{\ \ \ \ \ \ \nu}=0\, .
\end{equation}
Eq. (\ref{fermideltah2eq}) now takes the form of
\begin{equation}\label{deltah2final}
\Box \delta\bar{h}^{(2)}_{\mu\nu}+2R^{\eta\ \rho}_{\ \mu\ \nu}
\delta\bar{h}^{(2)}_{\eta\rho}=-2\delta S^F_{\mu\nu}
 \,.
\end{equation}
We define $\delta\bar{h}^{(2)}_{\mu\nu}$ to be the retarded solution 
of Eq. (\ref{deltah2final}) reading 
\begin{equation}\label{retdeltah2}
\delta\bar{h}^{(2)}_{\mu\nu}(x)
=\frac{1}{2\pi}\int
 G_{\mu\nu}^{\ \ \,\alpha'\beta'\, ret}[x|x']
\delta S^F_{\alpha'\beta'}(x')\sqrt{-g(x')}d^4x' \,.
\end{equation} 
Since the source term of Eq. (\ref{deltah2final}) diverges only like $r^{-2}$ the
integral in Eq. (\ref{retdeltah2}) has a finite contribution originating
from the vicinity of $z_G(\tau)$.
Notice that even though the expression 
in Eq. (\ref{retdeltah2}) satisfies Eq. (\ref{deltah2final}) 
 it is not a priori guaranteed that it also satisfies Eq. (\ref{fermideltah2eq}).
This equation will be satisfied only if the 
retarded solution (\ref{retdeltah2}) satisfies 
the Lorenz gauge conditions (\ref{lg}), this is shown in appendix B.

\subsection{General second-order solution} 

So far we have constructed a particular solution to Eq. (\ref{h2simple}), reading 
\begin{equation}\label{h2sofar}
\bar{h}^{(2)}_{\mu\nu}=\bar{\psi}^F_{\mu\nu}+\delta\bar{h}^{(2)}_{\mu\nu} \, .
\end{equation}
Having found one particular solution 
does not complete the construction, since we need to make sure 
that the constructed solution satisfies several required physical properties 
(e.g. it has to match the internal-zone 
solution). To find the desired physical solution 
we first construct the general solution to Eq. (\ref{h2simple}), and 
then impose a set of additional requirements on this solution. In this 
way we obtain a particular solution which is physically meaningful.

Since Eq. (\ref{h2simple}) is valid for $x\not\in z_G(\tau)$,
we find that we can construct a new solution by adding to $\bar{h}^{(2)}_{\mu\nu}$
a potential that satisfies a {\em semi-homogeneous} equation i.e.,
a homogeneous equation for $x\not\in z_G(\tau)$, reading
\begin{equation}\label{sheq} 
D_{\mu\nu}[\bar{h}^{(2)SH}]=0 \,\,\, ,x\not\in z_G(\tau)\,. 
\end{equation} 
The general solution to Eq. (\ref{h2simple}) is given by 
\begin{equation}\label{h2g}
\bar{h}^{(2)G}_{\mu\nu}\equiv \bar{h}^{(2)SH}_{\mu\nu}+\bar{h}^{(2)}_{\mu\nu}\, . 
\end{equation}
 where $\bar{h}^{(2)SH}_{\mu\nu}$ is the general solution to Eq. (\ref{sheq}).

\subsection{physical second-order solution}

To find a (particular) physically meaningful solution to Eq. (\ref{h2simple}) 
we need to impose additional requirements on $\bar{h}^{(2)G}_{\mu\nu}$.
These requirements can also be expressed 
as requirements imposed on 
the general semi-homogeneous solution $\bar{h}^{(2)SH}_{\mu\nu}$,
thus obtaining a particular semi-homogeneous solution.
For abbreviation we denote this particular 
semi-homogeneous solution with $\bar{\gamma}_{\mu\nu}$.
Using Eqs. (\ref{h2sofar},\ref{h2g}) the 
desired physical solution $\bar{h}^{(2)P}_{\mu\nu}$ is expressed as 
\begin{equation}\label{h2p}
\bar{h}^{(2)P}_{\mu\nu}=\bar{h}^{(2)}_{\mu\nu}+\bar{\gamma}_{\mu\nu}=
 \bar{\psi}^F_{\mu\nu}+\delta\bar{h}^{(2)}_{\mu\nu}+\bar{\gamma}_{\mu\nu}\,.
\end{equation}
We group the additional requirements into four groups 
(i) gauge conditions (ii) causality  requirements 
(iii) global boundary conditions (iv) boundary conditions at the worldline.

(i) First we impose gauge conditions. To obtain a simple representation for 
$\bar{\gamma}_{\mu\nu}$ we 
impose the Lorenz gauge conditions on $\bar{\gamma}_{\mu\nu}$. 
In this gauge Eq. (\ref{sheq}) takes the form of 
\begin{equation}\label{h2shpeq}
\Box \bar{\gamma}_{\mu\nu}+2R^{\eta\ \rho}_{\ \mu\ \nu}\bar{\gamma}_{\eta\rho}=0
 \,\,\,  ,x\not\in z_G(\tau)\,.
\end{equation}
Notice that by construction both $\delta\bar{h}^{(2)}_{\mu\nu}$ and
 $\bar{\gamma}_{\mu\nu}$ satisfy 
the Lorenz gauge conditions. 
However, $\bar{\psi}^F_{\mu\nu}$ 
does not satisfy these conditions,  
and therefore our particular second-order solution $\bar{h}^{(2)P}_{\mu\nu}$ 
in not in the (second-order) Lorenz gauge. 

(ii) Next we discuss causality. 
Causality is more easily discussed in terms of 
 an initial value formulation of the problem. 
Therefore, in this paragraph only we consider such an 
initial value formulation.
Suppose that we prescribe initial-data for the first-order and
second-order metric perturbations
on some initial spacelike hypersurface $\Sigma_0$, such that 
the corresponding constraint equations are satisfied 
on this hypersurface. 
Here we consider a standard extension of 
 the retarded solutions $\bar{h}^{(1)}_{\mu\nu}$, $\delta\bar{h}^{(2)}_{\mu\nu}$
[given by  Eq. (\ref{h1ret}), Eq. (\ref{retdeltah2})] to include
additional terms which depend on the initial-data.
Let $x$ be a point within the causal future $J^+(\Sigma_0)$. Then by construction 
the retarded solution $\bar{h}^{(1)}_{\mu\nu}(x)$ is unaffected 
by an arbitrary modification of the initial data on outside $J^-(x)\cap\Sigma_0$.
At second-order we define $\bar{\gamma}_{\mu\nu}$ 
to be the retarded solution of Eq. (\ref{h2shpeq}).  
Recall that $\delta\bar{h}^{(2)}_{\mu\nu}$ is the retarded solution 
of  Eq. (\ref{deltah2final}), and $\bar{\psi}^F_{\mu\nu}$
is completely determined from the first-order metric perturbations.
Eq. (\ref{h2p}) now implies that the particular solution 
$\bar{h}^{(2)P}_{\mu\nu}$ is unaffected by an arbitrary 
modification of the initial-data outside $J^-(x)\cap\Sigma_0$. 
In this sense the constructed second-order solution $\bar{h}^{(2)P}_{\mu\nu}(x)$ is 
manifestly causal.

(iii) In addition we require that the only source for 
$\bar{\gamma}_{\mu\nu}$ is the small black-hole. 
Since $\delta\bar{h}^{(2)}_{\mu\nu}$ satisfies an inhomogeneous equation 
(\ref{deltah2final}) it may contain waves that 
are not sourced by the worldline. 
However $\bar{\gamma}_{\mu\nu}$ satisfies a semi-homogeneous equation
(\ref{h2shpeq}), and therefore the 
waves within $\bar{\gamma}_{\mu\nu}$ can originate either from
 the worldline or from global boundary conditions 
(e.g. prescribed boundary conditions at $\cal{I}^-$). 
Since we are interested only in perturbations that are 
induced by the black-hole,   
we exclude any supplementary perturbations coming from 
these global boundary conditions.

(iv) We now turn to discuss the boundary conditions as $x \rightarrow z_G(\tau)$.
For this purpose let us consider once more a small 
but finite value of $\mu$. In this case  
Eq. (\ref{h2shpeq}) is valid only for $x\not\in S$. 
Following D'Eath analysis of first-order metric perturbations \cite{DEATH} we 
express a solution to this equation using a Kirchhoff representation. 
In this way $\bar{\gamma}_{\mu\nu}$ is expressed as a certain integral over 
a surface of a worldtube. Recall that the external-zone lies outside a worldtube 
with radius $r_E(\mu)$. Denoting the surface of this worldtube with 
$\Sigma_E$, we express $\bar{\gamma}_{\mu\nu}$ as 
\begin{equation}\label{kirchhoff}
\bar{\gamma}_{\mu\nu}(x)=-\frac{1}{4\pi}\int_{\Sigma_E(\mu)}\left(
G^{ret}_{\mu\nu\alpha'\beta'}[x|x']\nabla^{\epsilon'}\bar{\gamma}^{\alpha'\beta'}(x')-
\bar{\gamma}^{\alpha'\beta'}(x')\nabla^{\epsilon'}G^{ret}_{\mu\nu\alpha'\beta'}[x|x']
\right)d\Sigma_{\epsilon'}\,.
\end{equation}
Here $d\Sigma_{\epsilon'}$ denotes an outward directed three-surface element on 
$\Sigma_E$. In the derivation of Eq. (\ref{kirchhoff}) we 
assumed that $\bar{\gamma}_{\mu\nu}$ decays sufficiently fast at spatial infinity.
Furthermore, we assumed that the retarded Green's function
falls sufficiently fast into the past.
Consider substituting a given expansion of $\bar{\gamma}_{\mu\nu}$ (in powers of $r$)
into Eq. (\ref{kirchhoff}) and then taking the limit $\mu \rightarrow 0$.
Recall that at this limit $r_E \rightarrow 0$ and 
notice that $d\Sigma_\epsilon$ scales like $r_E^2$.
Therefore, only the diverging terms (as $r\rightarrow 0$) in this expansion 
 give rise to a non-vanishing contribution to 
$\bar{\gamma}_{\mu\nu}$ at the limit $\mu\rightarrow 0$. 
We conclude that at the limit, it is sufficient to specify 
{\em divergent boundary conditions} to obtain 
a unique physical solution to Eq. (\ref{h2shpeq}).
To obtain these boundary conditions 
we examine the divergent behavior 
of  $\bar{\gamma}_{\mu\nu}(x)$ as $x\rightarrow z_G(\tau)$. For
this  we use 
Eq. (\ref{h2p}) together with an analysis of the   
behavior of $\bar{\psi}^F_{\mu\nu}$, $\delta\bar{h}^{(2)}_{\mu\nu}$, 
and $\bar{h}^{(2)P}_{\mu\nu}$ near $r=0$. 

First let us consider the divergent behavior of $\bar{\psi}^F_{\mu\nu}$. 
Using Eqs. 
(\ref{h1decomp},\ref{h1s},\ref{1fermigauge}) together with Eqs. (\ref{quadratic},\ref{psiexplicit}) 
we obtain the following expansion for $\bar{\psi}_{\mu\nu}^F$ 
in Fermi normal coordinates 
\begin{equation}\label{psiexpan}
\bar{\psi}_{\mu\nu}^F\stackrel{*}=-\frac{1}{4r^2}[2u_{\mu}u_{\nu}+7\eta_{\mu\nu}]+O(r^0)\,.
\end{equation}
Next we examine the behavior of $\delta\bar{h}^{(2)}_{\mu\nu}$ near 
$r=0$. Solving Eq. (\ref{deltah2final}) iteratively (see Appendix C) 
shows that $\delta\bar{h}^{(2)}_{\mu\nu}$ is bounded as $r\rightarrow 0$. 
Finally, we have to determine the divergent behavior of $\bar{h}^{(2)P}_{\mu\nu}$
in the vicinity of $r=0$. This requires an analysis of 
the internal-zone solution, which is discussed in the next section. 

\section {Approximation in the internal-zone}\label{internal}

To simplify the discussion we assume that all 
the length scales characterizing
the background spacetime $\{{\cal R}_i\}$ are of the same order of magnitude 
$\cal{R}$. Recall that $\mu\ll \cal{R}$, and 
 furthermore in the internal-zone we have $r\ll \cal{R}$. 
By virtue of the smallness of $r/{\cal R}$ and $\mu/\cal{R}$ 
we may expand the full spacetime metric in the internal-zone as follows
\begin{equation}\label{internalexp}
g_{\mu\nu}=g^{Sch}_{\mu\nu}
+{\cal R}^{-1}g^{(1)}_{\mu\nu}+{\cal R}^{-2}g^{(2)}_{\mu\nu}
+O({\cal R}^{-3})\,.
\end{equation}
Here $g^{Sch}_{\mu\nu}$ is the metric of the small Schwarzschild black-hole.
Recall that in the buffer-zone both expansions (\ref{metricexp}) 
and (\ref{internalexp}) are valid, and therefore we can 
formally expand these two equations simultaneously using the fact that 
in the buffer-zone both $\mu r^{-1}$ and $r{\cal R}^{-1}$ are small. 
Following Thorne and Hartle \cite{TH} the various 
dimensional combinations involved in these expressions can be 
summarized in a table (see Table I).
\begin{table*}[h]
 \begin{flushleft}
\caption{Schematic representation of dimensional quantities combinations in 
expansions of the metric in the buffer zone.
The top row gives the metric's external-zone expansion (\ref{metricexp}), and the left column 
gives the metric's internal-zone expansion (\ref{internalexp}).}
\begin{tabular*}{0.50\textwidth}{@{\extracolsep{\fill}}l||l|l|l|l}
$ $ &  $g_{\mu\nu}$ &  ${\mu}h^{(1)}_{\mu\nu}$ &  ${\mu^2}h^{(2)}_{\mu\nu}$ & $...$\\
\hline
\hline
$g^{Sch}_{\mu\nu}$  &  $\eta$    &  $\mu r^{-1}$&  ${\mu^2}r^{-2}$ &  $...$\\
\hline
${\cal R}^{-1}g^{(1)}_{\mu\nu}$  & $r{\cal R}^{-1}$    &  $\mu{\cal R}^{-1}$ &
${\mu^2}{(r{\cal R})}^{-1}$ &  $...$\\
\hline
${\cal R}^{-2}g^{(2)}_{\mu\nu}$  & $r^2{\cal R}^{-2}$    &  $\mu r {\cal R}^{-2}$ &
  ${\mu^2}{\cal R}^{-2}$ &  $...$\\
\hline
$...$  & $...$    &  $...$  &  $...$ &  $...$\\

\end{tabular*}
\end{flushleft}
\end{table*} 

The top row in Table I gives the external-zone expansion, and the left most column 
in this table gives the internal-zone expansion. 
In the buffer-zone where both expansions 
are valid one can find appropriate coordinates in which these two expansions coincide. 
Each entry in this table schematically represents the combinations of the 
 dimensional quantities ($\mu$,$\cal{R}$, and $r$) obtained 
from the simultaneous expansions of the top row and left most column. 
Note that this table provides only the powers of the 
relevant dimensional combinations and does not give the exact expressions.

As before the background metric $g_{\mu\nu}$ is 
described in Fermi normal coordinates based on $z_G(\tau)$.
The Schwarzschild metric $g^{Sch}_{\mu\nu}$ is described in the 
Schwarzschild isotropic coordinates (\ref{isotropic}). 
$\eta$ (schematically) denotes the Minkowski metric, which is the leading 
order is both the expansions of $g_{\mu\nu}$ and $g^{Sch}_{\mu\nu}$ 
in the buffer-zone.

We number the rows from top to bottom 
starting at the row of $g^{Sch}_{\mu\nu}$ (row $0$), and number the columns 
from left to right starting from the column of $g_{\mu\nu}$ (column $0$). 
Each entry in the table is associated with an ordered pair of numbers: (row,column).
The divergent behavior of $\bar{h}^{(2)P}_{\mu\nu}$ near the worldline
follows from column $2$. 
In this column only terms $(0,2)$ and $(1,2)$ need to be considered 
since only these terms have the potential of producing 
divergent terms as $r\rightarrow 0$. 
These terms follow from an expansion for $g^{Sch}_{\mu\nu}$ and 
${\cal R}^{-1}g^{(1)}_{\mu\nu}$ that we discuss next. 

Let us consider first row $0$. In the Schwarzschild isotropic coordinates
 (\ref{isotropic}) the entries  $(0,1)$ and $(0,2)$ are nothing but the 
terms $\mu H^{(1)}_{\mu\nu}$ and $\mu^2 H^{(2)}_{\mu\nu}$ 
[see Eq.(\ref{schexp})], respectively; and
their explicit form can be easily obtained from Eq. (\ref{bigh}). 
We identify $\tilde{r}$ with $r$, with this we find that 
the term $(0,1)$ coincide with the corresponding terms 
of order $r^{-1}$ in the expansion of 
$\mu h^{(1)}_{\mu\nu}$.

We now discuss row $1$. Expanding the background metric in the vicinity 
of the worldline gives  $g_{\mu\nu}=\eta_{\mu\nu}+O(r^2{\cal R}^{-2})$,
 which implies that the term $(1,0)$ vanishes. 
This vanishing term serves as a boundary condition for
 the perturbations equations for ${\cal R}^{-1}g^{(1)}$ as $\tilde{r}\rightarrow \infty$.
The term  ${\cal R}^{-1}g^{(1)}$ is obtained by solving 
the a gravitational vacuum perturbations equation in a Schwarzschild background, with 
these boundary conditions.
However, it is well known that these $O({\cal R}^{-1})$ perturbations can be 
eliminated by a gauge choice and mass redefinition.
Therefore, we may always choose a gauge in which row 1 vanishes identically 
(see also \cite{TH}). 
Notice that the vanishing of the $(1,1)$ term 
conforms with the fact that the $O(r^0)$ terms are absent from the expansion of 
$\mu h^{(1)}_{\mu\nu}$ in Fermi gauge.
Since the $(1,2)$ term vanishes we conclude that the only divergent term 
in the expansion of $\bar{h}^{(2)P}_{\mu\nu}$ near the worldline, is a term 
which diverges like $r^{-2}$. The form of this term  
is provided by $\bar{H}^{(2)}_{\mu\nu}$ in Eq. (\ref{bigh}).
Notice that $\bar{H}^{(2)}_{\mu\nu}$ is identical to the divergent expression
of $\bar{\psi}_{\mu\nu}^F$ given by Eq. (\ref{psiexpan}).

\section{Conclusions}\label{conclusion}

We have found that the divergent terms in the 
expansion for  $\bar{h}^{(2)P}_{\mu\nu}$
 coincides with the divergent terms in 
the expansion for $\bar{\psi}^F_{\mu\nu}$ given by Eq. (\ref{psiexpan}).
Rewriting Eq. (\ref{h2p}) as
\[
\bar{\gamma}_{\mu\nu}=\bar{h}^{(2)P}_{\mu\nu}
- \bar{\psi}^F_{\mu\nu}-\delta\bar{h}^{(2)}_{\mu\nu}\,,
\]
and recalling that $\delta\bar{h}^{(2)}_{\mu\nu}$ is bounded as $r\rightarrow 0$,
we find that at this limit $\bar{\gamma}_{\mu\nu}$ is bounded as well. 
Recall that only divergent boundary conditions as $r \rightarrow 0$
can produce a non-vanishing semi-homogeneous retarded solution $\bar{\gamma}_{\mu\nu}$.
Since the divergent boundary conditions of Eq. (\ref{h2shpeq}) vanish, 
we find that $\bar{\gamma}_{\mu\nu}$ vanishes identically.

From Eq. (\ref{h2p}) we finally conclude that the physical second-order 
 gravitational perturbations in the external-zone are given by 
\begin{equation}\label{finalconc}
\bar{h}^{(2)P}_{\mu\nu}=\bar{\psi}^F_{\mu\nu}+\delta\bar{h}^{(2)}_{\mu\nu}\,.
\end{equation}
Here $\bar{\psi}^F_{\mu\nu}$ is given by 
Eq. (\ref{psiexplicit}), where the perturbations $\bar{h}^{(1)}_{\mu\nu}$
are in Fermi gauge, and 
$\delta\bar{h}^{(2)}_{\mu\nu}$ is given by Eq. (\ref{retdeltah2}). 
Eq. (\ref{finalconc}) provides a simple covariant prescription for the construction of 
the second-order metric perturbations without 
any reference to a particular (background) coordinate system.

\acknowledgments

I am grateful to Amos Ori and to Eric Poisson 
for numerous valuable discussions.
This work was supported in part by the Natural Sciences and Engineering
Research Council of Canada, and also in part by 
The Israel Science Foundation (grant no. 74/02-11.1).
\appendix 
\section{Expansion of Ricci tensor}
The linear and quadratic terms in the expansion of Ricci tensor 
[see Eq.(\ref{riccidec})] are given by (see e.g. \cite{MST})
\begin{eqnarray}
\bar{R}^{(L)}_{\mu\nu}[h]&\equiv&D_{\mu\nu}[\bar{h}]\equiv
\frac{1}{2}\left[-\bar{h}^{\ \ \ \ \,\alpha}_{\mu\nu;\alpha}+
\bar{h}^{\ \ \ \ \,\alpha}_{\alpha\mu;\nu}+
\bar{h}^{\ \ \ \ \,\alpha}_{\alpha\nu;\mu}-
g_{\mu\nu}\bar{h}^{\ \ \ \ \,\beta\alpha}_{\beta\alpha;}
\right]\,,\\
{R}^{(Q)}_{\mu\nu}[h]&\equiv&\frac{1}{2}
\Biglb[\frac{1}{2}{h}_{\alpha\beta;\mu}{h}^{\alpha\beta}_{\ \ ;\nu}
+{h}^{\alpha\beta}({h}_{\alpha\beta;\mu\nu}+{h}_{\mu\nu;\alpha\beta}
-2{h}_{\alpha(\mu;\nu)\beta})\\\nonumber
&&+2{h}_{\nu}^{\ \alpha;\beta}{h}_{\mu[\alpha;\beta]}
-({h}^{\alpha\beta}_{\ \ \,;\beta}-\frac{1}{2}{h}^{;\alpha})
(2{h}_{\alpha(\mu;\nu)}-{h}_{\mu\nu;\alpha})
\Bigrb]\,.
\end{eqnarray}
We also used the notation ${S}_{\mu\nu}[\bar{h}]\equiv-\bar{R}_{\mu\nu}^{(Q)}[h]$,
where on the right hand side $h_{\mu\nu}$ is expressed with
$\bar{h}_{\mu\nu}$.
\section{$\delta\bar{h}^{(2)}_{\mu\nu}$ satisfies Lorenz gauge conditions}

Here we show that the Lorenz gauge conditions are indeed satisfied 
by the retarded solution (\ref{retdeltah2}).
For this purpose we follow a standard method of deriving differential equations for 
$\nabla^\nu \delta\bar{h}^{(2)}_{\mu\nu}$.
By applying divergence operator
to Eq. (\ref{deltah2final}) and using a contraction of Bianchi identities  
together with the fact that background geometry is a vacuum spacetime,
we obtain
\begin{equation}\label{divdeltah2final}
\Box(\nabla^\nu \delta\bar{h}^{(2)}_{\mu\nu})
=-2\nabla^\nu \delta S^F_{\mu\nu}\,.
\end{equation}
We assume that Lorenz gauge conditions (\ref{lg}) are satisfied on an initial 
spacelike hypersurface $\Sigma_I$ and moreover that 
$[(n^\alpha\nabla_\alpha)\nabla^\nu \delta\bar{h}^{(2)}_{\mu\nu}]_{\Sigma_I}=0$ where 
$n^\alpha$ is normal to $\Sigma_I$. We shall now show that
the retarded solution of Eq. (\ref{divdeltah2final}) vanishes, and therefore 
$\delta\bar{h}^{(2)}_{\mu\nu}$ satisfies the Lorenz gauge conditions as required.

First, we will show that the source of Eq. (\ref{divdeltah2final}) vanishes
 for $x\not\in z_G(\tau)$.
Consider a metric $\hat{g}_{\mu\nu}$ that 
depends on a small parameter $\mu$, and may be expanded as follows
\begin{equation}\label{formalexp}
\hat{g}_{\mu\nu}(x)=g_{\mu\nu}(x)+
\mu g^{(1)}_{\mu\nu}(x)+\mu^2 g^{(2)}_{\mu\nu}(x)+O(\mu^3)\,.   
\end{equation}
Here $\hat{g}_{\mu\nu}(x)$ is not 
necessarily a solution of Einstein's field equations in vacuum, whereas $g_{\mu\nu}$
maintain its definition as a vacuum solution to Einstein's field equations.
We now employ Bianchi identities reading
\begin{equation}\label{Bianchi}
\hat{g}^{\alpha\mu}\hat{\nabla}_\alpha \hat{G}_{\mu\nu}\equiv 0\,.
\end{equation}
Here the contravariant metric satisfies 
$\hat{g}^{\alpha\mu}\hat{g}_{\mu\beta}=\delta^\alpha_\beta$, 
 $\hat{\nabla}_\mu$ denotes the covariant derivative with respect to 
$\hat{g}_{\mu\nu}$, and $\hat{G}_{\mu\nu}$ is Einstein tensor 
evaluated with this metric. 
We now employ decomposition (\ref{formalexp}) to
formally expand Einstein tensor, and the
covariant derivative (for rank-2 tensors), giving
\begin{equation}\label{Einsteindec}
\hat{G}_{\mu\nu}=G_{\mu\nu}^{(0)}+\mu G_{\mu\nu}^{(1)}+
\mu^2 G_{\mu\nu}^{(2)}+O(\mu^3)\,.
\end {equation}
\begin{equation}\label{gradexp}
\hat{g}^{\alpha\mu}\hat{\nabla}_{\alpha}=\nabla^{\mu}+
\mu {\Gamma}_{1}^\mu+\mu^2 {\Gamma}_{2}^\mu
+O(\mu^3)\,.
\end {equation}
In these expansions the dependence on $\mu$ is only through the explicit 
powers $\mu^{i}$, ${\Gamma}_{1}^\mu$ and  ${\Gamma}_{2}^\mu$ denote 
linear operators (defined on rank-2 tensors), 
whose explicit form is not required here.
We now substitute Eqs. (\ref{Einsteindec},\ref{gradexp}) into 
(\ref{Bianchi}) and obtain a perturbative expansion of Bianchi identities. 
The Bianchi identities are valid for any value of $\mu$ and therefore 
the individual terms in their expansion in powers of $\mu$ vanish 
identically, yielding the following set of identities for an arbitrary tensor fields
$g^{(1)}_{\mu\nu}$ and $g^{(2)}_{\mu\nu}$
\begin{eqnarray}\label{1stBianchi}
&&\nabla^{\mu}D_{\mu\nu}[\bar{g}^{(1)}]\equiv 0\,,\\
\label{2ndBianchi}
&&\nabla^{\mu}G^{(2)}_{\mu\nu}+ {\Gamma}_{1}^\mu\biglb[D_{\alpha\beta}[\bar{g}^{(1)}]\bigrb]\equiv0
\,.
\end{eqnarray}
Here we denoted
\begin{equation}\label{g2}
G^{(2)}_{\mu\nu}\equiv D_{\mu\nu}[\bar{g}_2]-S_{\mu\nu}[\bar{g}_1]
+\frac{1}{2}R^{(L)}_{\alpha\beta}[g_1](g^{(1)\alpha\beta}g_{\mu\nu}
-g^{(1)}_{\mu\nu}g^{\alpha\beta})\,.
\end{equation}
Employing Eqs. (\ref{finalfst},\ref{1stBianchi},\ref{2ndBianchi},\ref{g2})
 we find that 
the source term of Eq. (\ref{divdeltah2final}) vanishes for
$x\not\in z_G(\tau)$. 

Next, we show that the source of Eq. (\ref{divdeltah2final}) on 
the worldline is too weak to produce a non-vanishing  
$\nabla^{\nu}\delta\bar{h}^{(2)}_{\mu\nu}$.
We shall now estimate the strength of the source Eq. (\ref{divdeltah2final}) 
on the worldline. Consider a hypersurface of constant time, generated 
by spacelike geodesics which are normal to $z_G(\tau)$. 
In this hypersurface we consider a small sphere $D(\epsilon)$ of radius $\epsilon$,
 centered at $r=0$; and calculate 
the following three dimensional volume integral over $\nabla^{\nu}\delta S_{\mu\nu}$ 
inside this sphere, reading
\begin{equation}\label{intdivdeltas}
\int_{D(\epsilon)}\bar{g}^{\mu'}_{\mu}(z_G,x')\nabla^{\nu'}\delta S^F_{\mu'\nu'} dV'\,. 
\end{equation} 
If this integral vanishes then the strength of 
the source term on the world line is weaker 
than a delta-function source term, and it is too weak
to produce a non-vanishing $\nabla^{\nu}\delta\bar{h}^{(2)}_{\mu\nu}$.
Recall that $\nabla^{\nu}\delta S^F_{\mu\nu}$ vanishes for $r\not =0$.
Therefore, we may take the limit $\epsilon \rightarrow 0$ without changing 
the value of this integral. Explicit expression of  
$\nabla_{\nu}\delta S_{\mu}^{F\,\nu}$ reads
\begin{equation}\label{divexp}
\nabla_{\nu}\delta S_{\mu}^{F\,\nu}=
{(-g)}^{-1/2}\frac{\partial}{\partial x^a}(\delta S_{\mu}^{F\,a}\sqrt{-g})+
{(-g)}^{-1/2}\frac{\partial}{\partial x^0}(\delta S_{\mu}^{F\,0}\sqrt{-g})-
\frac{1}{2}g_{\nu\rho,\mu}\delta S^{F\nu\rho}\,.
\end{equation}
We now substitute Eq. (\ref{divexp}) into Eq. (\ref{intdivdeltas}), and evaluate
this integral using Fermi normal coordinates, based on the worldline.
Notice that in these coordinates $\bar{g}^{\mu'}_{\mu}(z_G,x')=\delta_{\mu}^{\mu'}+O(r^2)$,
$dV$ scales like $r^2$, while the second and third terms in 
Eq. (\ref{divexp}) scale like $r^{-2}$ and $r^{-1}$, respectively. We therefore find that
at the limit  $\epsilon \rightarrow 0$ the integral (\ref{intdivdeltas}) over 
the second and third terms in Eq. (\ref{divexp}) vanishes. Substituting the  
first term in Eq. (\ref{divexp}) into integral (\ref{intdivdeltas}) and using 
Gauss theorem we find that at for small values of $\epsilon$ 
the integral (\ref{intdivdeltas}) is approximated by 
\begin{equation}
\oint_{\partial D(\epsilon)} \delta S_{\mu}^{F\,a'} d \Sigma_{a'}\,.
\end{equation} 
Here $\partial D(\epsilon)$ is the surface of the sphere. 
Consider an expansion of $\delta S_{\mu}^{F\,a'}$  in powers of $r$  
in the vicinity of $r=0$. Here only terms which 
scale like $r^{-2}$ have the potential of producing a non-vanishing 
integral at the limit $\epsilon\rightarrow 0$.
Using the schematic form (\ref{deltah2source}) one finds that 
only terms of the form $\bar{h}^{(1)S}\nabla \nabla \bar{h}^{(1)S}$ and 
 $\nabla\bar{h}^{(1)S}\nabla \bar{h}^{(1)S}$ produce 
terms in $\delta S_{\mu}^{F\,a'}$  which scale like $r^{-2}$.  
We now consider an  expansion of $\bar{h}^{(1)S}$ and its derivatives, see 
 Eqs. (\ref{h1s},\ref{gradh1s},\ref{2gradh1s}) for 
the leading terms in these expansions.
Examining these equations we see that these expansions depends 
on dimensionless quantities of the form $u^\mu,\Omega^\mu,\eta_{\mu\nu}$, 
higher order terms in these expansions also include the Riemann tensor and
its derivatives. 
Dimensional analysis implies that the 
terms in $\delta S_{\mu}^{F\,a'}$  which scale like $r^{-2}$
must be linear in Riemann tensor . 
The integral in Eq. (\ref{intdivdeltas}) provides us with a vector at $r=0$.
When integrating over the above mentioned local expansions we find that 
this vector must be composed of Riemann tensor, $u^\mu$, and the background metric. 
However, in a vacuum background spacetime
one can not construct form these quantities a non-vanishing vector. Therefore the 
integral in Eq. (\ref{intdivdeltas}) vanishes. 

In the above calculations we showed that
the source of Eq. (\ref{divdeltah2final}) vanishes
 for $x\not\in z_G(\tau)$, and furthermore that 
a volume integral over this source, which includes the worldline
vanishes as well. We therefore find (with the above mentioned initial conditions)
that the retarded solution to of Eq. (\ref{divdeltah2final}) vanishes, and therefore 
$\delta\bar{h}^{(2)}_{\mu\nu}$ satisfies the Lorenz gauge conditions.

\section{$\delta\bar{h}^{(2)}_{\mu\nu}$ is bounded  as $r\rightarrow 0$}

We show that the retarded potential 
$\delta\bar{h}^{(2)}_{\mu\nu}$ given by
Eq. (\ref{retdeltah2}) is bounded as $r\rightarrow 0$. 
Recall that the source term in Eq. (\ref{deltah2final}) 
diverges like $r^{-2}$ as $r\rightarrow 0$. Therefore, 
the integral in Eq. (\ref{retdeltah2}) converges for 
 $x\not\in z_G(\tau)$.
In particular $\delta\bar{h}^{(2)}_{\mu\nu}$ is finite on the surface 
of a worldtube which surrounds the worldline at a fixed spatial distance $r=r_B$,
where $r_B<<{\cal R}$.
We now consider the solution of  Eq. (\ref{deltah2final}) within this
worldtube. By virtue 
of the smallness of $r{\cal R}^{-1}$ within this worldtube, Eq. (\ref{deltah2final})
 can be solved iteratively using the following
expansions of  $\delta\bar{h}^{(2)}_{\mu\nu}$ and $g_{\mu\nu}$
\begin{equation}\label{deltah2exp}
\delta\bar{h}^{(2)}_{\mu\nu}=\delta\bar{h}^{(2)}_{(0)\mu\nu}+
{\cal R}^{-1}\delta\bar{h}^{(2)}_{(1)\mu\nu}+
{\cal R}^{-2}\delta\bar{h}^{(2)}_{(2)\mu\nu}+O({\cal R}^{-3})\,,
\end{equation}
\begin{equation}
g_{\mu\nu}\stackrel{*}=\eta_{\mu\nu}+O({\cal R}^{-2})\,.
\end{equation}
Here again we employ Fermi normal coordinates based on $z_G(\tau)$.
Note that the time scale in which the source term of Eq. (\ref{deltah2final})
changes is of $O({\cal R})$ and therefore the
leading term $\delta\bar{h}^{(2)}_{(0)\mu\nu}$ satisfies the following equation
\begin{equation}\label{deltah2sub0}
(\delta^{ab}\partial_a\partial_b) 
\delta\bar{h}^{(2)}_{(0)\mu\nu}\stackrel{*}=-2 \delta S^F_{\mu\nu}\,.
\end{equation}
Here $x^a,x^b$ denote the spatial Fermi normal coordinates. 
Eq. (\ref{deltah2sub0}) is a set of Poisson's equations 
for each tensorial component of $\delta\bar{h}^{(2)}_{(0)\mu\nu}$. 
To solve these equations we decompose into spherical harmonics centered at $r=0$
, reading
\[
\delta\bar{h}^{(2)}_{(0)\mu\nu}\stackrel{*}=\sum_{lm}Y^{lm}\phi^{lm}_{\mu\nu}\,,
\]
\[
-2 \delta S^F_{\mu\nu} \stackrel{*}=\sum_{lm}Y^{lm}\rho^{lm}_{\mu\nu}\,.
\]
The solution for each spherical harmonics component is given by
\begin{equation}\label{philm}
\phi^{lm}_{\mu\nu}(r)\stackrel{*}=-\frac{1}{2l+1}\int_{0}^{r_
B}
\frac{r_<^l}{r_>^{l+1}}{r'}^2\rho^{lm}_{\mu\nu}(r')dr'+B.T. \ \ .
\end{equation}
Here $r_{>}$ and $r_<$ are the larger and smaller terms from the pair $\{r,r'\}$,
respectively; $B.T.$ denotes finite boundary terms coming from the contribution 
of the surface of the worldtube. Expanding $\rho^{lm}_{\mu\nu}$ in a power series
gives
\[
\rho^{lm}_{\mu\nu}(r)\stackrel{*}=a^{lm}_{\mu\nu(-2)}r^{-2}+a^{lm}_{\mu\nu(-1)}r^{-1}+
O(r^{0})\,.
\]
Eq. (\ref{philm}) implies that in the expansion of $\rho^{lm}_{\mu\nu}$ 
only the term $a^{00}_{\mu\nu(-2)}r^{-2}$ gives rise to a (logarithmic) divergency in
$\phi^{lm}_{\mu\nu}$, while all the other terms produce a bounded potential at $r=0$. 
Using the schematic expression (\ref{deltah2source}) one finds that 
the terms in the source of Eq. (\ref{deltah2final})
that can possibly contribute to the problematic term $a^{00}_{\mu\nu(-2)}r^{-2}$, 
are of the form $\bar{h}^{(1)S}\nabla \nabla \bar{h}^{(1)S}$ and 
 $\nabla\bar{h}^{(1)S}\nabla \bar{h}^{(1)S}$.
These terms can be expanded in the vicinity $z_G(\tau)$ using expansions
 (\ref{h1s},\ref{gradh1s},\ref{2gradh1s}) evaluated to a higher accuracy (see discussion at the end of
Appendix B). Dimensional analysis
implies that $a^{00}_{\mu\nu(-2)}$ has to be proportional to a component 
of Riemann tensor. Since we assumed a vacuum 
background spacetime the only possible non-vanishing candidate for $a^{00}_{\mu\nu(-2)}$
is $R_{\mu\alpha\nu\beta}u^{\alpha}u^{\beta}$ times a constant.
Explicit calculation (using MATHEMATICA software) of the constant coefficient of this $l=0$ term 
shows that it vanishes, which implies that $\delta\bar{h}^{(2)}_{(0)\mu\nu}$ is bounded at
$r=0$.
The higher order corrections to $\delta\bar{h}^{(2)}_{(0)\mu\nu}$
 given by Eq. (\ref{deltah2exp}) are smaller than the leading term 
by at least a factor of $r{\cal R}^{-1}$, and are therefore bounded as well. 
We conclude that $\delta\bar{h}^{(2)}_{\mu\nu}$ is bounded as $r \rightarrow 0$.



\begin{thebibliography}{99}
\bibitem{sfgauge}
To verify this statement one can simply substitute the expression 
for  $h^{(1)R(new)}_{\mu\nu}$ into  Eq. (\ref{selfaccel}).
The difference between the new self-force expression and 
the original Lorenz gauge self-force conforms with the expression 
for the self-force gauge transformation derived by Barack and Ori \cite{BOGAUGE}.
\bibitem{nullcont}
To construct this continuation consider the
 future null-cones $\Sigma_\tau$ emanating from the worldline $z_G(\tau)$. 
Here we focus on a local neighborhood of the worldline in which
these null-cones do not intersect each other.
For an arbitrary point $z_G({\tau^-})$ 
one may choose an arbitrary continuation of $\xi^\mu$ on $\Sigma_{\tau^-}$,
such that $\xi^\mu$ decays to zero away from the worldline.
In this way the constructed gauge preserve causality in the following sense:
The perturbations on the null-cones $\Sigma_{\tau}$ for $\tau\le \tau^-$ will
 remain unchanged if one modifies the worldline for 
$\tau>\tau^-$. Such a modification of the worldline 
is possible by introducing additional GW that 
interact with the worldline for $\tau>\tau^-$.
\end{thebibliography}
\end{document}